# Floating Cities, Islands and States*

**Alexander Bolonkin**
C&R, 1310 Avenue R, #F-6, Brooklyn, NY 11229, USA
T/F 718-339-4563, aBolonkin@juno.com, aBolonkin@gmail.com, http://Bolonkin.narod.ru

## Abstract

Many small countries are in need of additional territory. They build landfills and expensive artificial islands. The ocean covers 71% of the Earth's surface. Those countries (or persons of wealth) starting the early colonization of the ocean may obtain advantages through additional territory or creating their own independent state. An old idea is building a big ship. The best solution to this problem, however, is the provision of floating cities, islands, and states. The author's idea is to use for floating cities, islands, and states a cheap floating platform created from a natural ice field taken from the Arctic or Antarctic oceans. These cheap platforms protected by air-film (bottom and sides) and a conventional insulating cover (top) and having a cooling system can exist for an unlimited time. They can be increased in number or size at any time, float in warm oceans, travel to different continents and countries, serve as artificial airports, harbors and other marine improvements, as well as floating cities and industrial bases for virtually any use.

Author researches and computes parameters of these ice floating platforms, other methods of building such floating territory, compares them and shows that the offered method is the most cheap and efficient means of ocean colonization.

-------------------------------
*Presented in http:arxiv.org

## Introduction

**Short information about oceans, history large ship, and ice fields.**

An **ocean** is a major body of saline water, and a principal component of the hydrosphere. Approximately 71% of the Earth's surface (an area of some 361 million square kilometers) is covered by ocean, a continuous body of water that is customarily divided into several principal oceans and smaller seas. More than half of this area is over 3,000 meters (9,800 ft) deep. Average oceanic salinity is around 35 parts per thousand (ppt) (3.5%), and nearly all seawater has a salinity in the range of 31 to 38 ppt.

The area of the World Ocean is 361 million square kilometers (139 million sq mi), its volume is approximately 1.3 billion cubic kilometers (310 million cu mi), and its average depth is 3,790 meters (12,430 ft). Nearly half of the world's marine waters are over 3,000 meters (9,800 ft) deep. The vast expanses of deep ocean (anything below 200m) cover about 66% of the Earth's surface. This does not include seas not connected to the World Ocean, such as the Caspian Sea.

The total mass of the hydrosphere is about $1.4 \times 10^{21}$ kilograms, which is about 0.023% of the Earth's total mass. Less than 2% is freshwater, the rest is saltwater, mostly in the ocean.

Though generally recognized as several 'separate' oceans, these waters comprise one global, interconnected body of salt water often referred to as the World Ocean or global ocean.
That includes: Pacific Ocean, the Atlantic Ocean, the Indian Ocean, the Southern Ocean, and the Arctic Ocean.

**Ocean colonization** is the theory and practice of permanent human settlement of oceans. Such settlements may float on the surface of the water, or be secured to the ocean floor, or exist in an intermediate position.

One primary advantage of ocean colonization is the expansion of livable area. Additionally, it might offer various other possible benefits such as expanded resource access, novel forms of governance (for instance micronations), and new recreational activities.

Many lessons learned from ocean colonization will likely prove applicable to space colonization. The ocean may prove simpler to colonize than space and thus occur first, providing a proving ground for the latter. In particular, the issue of sovereignty may bear many similarities between ocean and space colonization; adjustments to social life under harsher circumstances would apply similarly to the ocean and to space; and many technologies may have uses in both environments.

**Economy of ocean.** Central to any practical attempt at ocean colonization will be the underlying economic reality. To become self-sustaining, the colony will aim to produce output of a kind which holds a comparative advantage by occurring on the ocean. While it can save the cost of acquiring land, building a floating structure that survives in the open ocean has its own costs. Oceanfront land can hold a very high value, especially in countries with no income taxes, so building space and selling it may prove popular. Tourists often visit warm locales during the winter; indeed, tourism drives the economies of many small island nations. The colony might also compete as an offshore financial centre.

While importing food and fishing may compose the majority of ocean settlement food consumption, other possibilities include hydroponics and open-ocean aquaculture. Thus, an ocean settlement may be either a net importer or a net exporter of food goods.

Such settlements or cities would probably import diesel and run conventional power plants as small islands everywhere do. However, other possibilities include solar power, nuclear plants, deep-sea oil deposits, and developing/farming a species of seaweed or algae for biofuel.

Ocean thermal energy conversion (OTEC) is another potential energy source. All that is required is tropical (warm) surface water and access to deep, very cold water. The difference in temperature is used to drive an electric generator via a turbine. (There is an added benefit in that the deep cold water usually is more fertile than surface water in the open ocean, and can support mariculture).

Similar communities already exist in the form of hotels, research stations, houseboats, houses on stilts, land below sea level behind dikes, vacation cruise ships, ocean oil rigs, etc.

Further, humans spreading to small islands throughout the world has already occurred and is ongoing. Using current technology to create artificial islands is just an incremental step in continuing the spread of humanity.

**Millennial projects.** An artificial (ground, modeless) island is an island that has been constructed by humans rather than formed by natural means. They are created by expanding existing islets, construction on existing reefs, or amalgamating several natural islets into a bigger island.

An internet mailing list formed to attempt to organize it. The group incorporated as the "Living Universe Foundation." The list was still in existence as of 2007.

Some contemporary projects are much more ambitious. Kansai International Airport is the first airport to be built completely on an artificial island in 1994, followed by Chūbu Centrair International Airport in 2005 and the New Kitakyushu Airport and Kobe Airport in 2006.

Dubai is home to some of the largest artificial island complexes in the world, including the three Palm Islands projects, The World and the Dubai Waterfront, the last of which will be the largest in scale.

The Israeli government is now planning for 4 artificial islands to be completed in 2013, off the coasts of Tel Aviv, Herzliya, Netanya and Haifa. Each island will house some 20,000 people and bring in 10,000 jobs. The islands should help with overcrowding in Israeli cities and even be employed to do the same in Gaza.

A proposal has also been presented in The Netherlands to create artificial islands, perhaps in the shape of a tulip, in the North Sea.

Under the United Nations Convention on the Law of the Sea treaty (UNCLOS), artificial islands have little legal recognition. Such islands are not considered harbor works (Article 11) and are under the jurisdiction of the nearest coastal state if within 200 nautical miles (370 km) (Article 56). Artificial islands are not considered islands for purposes of having their own territorial waters or exclusive economic zones, and only the coastal state may authorize their construction (Article 60). However, on the high seas beyond national jurisdiction, any "state" may construct artificial islands (Article 87).

Some attempts to create micronations have involved artificial islands such as Sealand and Republic of Rose Island.

**Big ship projects.**

**America World City** (originally named Phoenix World City) is a concept for a floating city proposed by John Rogers of World City Corporation. It is conceived as the first of three such behemoths serving United States ports and flying the U. S. Flag. Rogers died in October, 2005.

**Freedom Ship.** Freedom Ship was a concept proposed by Norman Nixon. One has a design length of 4,500 feet (1400 m), a width of 750 feet (230 m), and a height of 350 feet (110 m), *Freedom Ship* would be more than 4 times longer than the *Queen Mary*. The design concepts include a mobile modern city featuring luxurious living, an extensive duty-free international shopping mall, and a full 1.7 million square foot (160,000 m²) floor set aside for various companies to showcase their products. *Freedom Ship* would not be a cruise ship, it is proposed to be a unique place to live, work, retire, vacation, or visit. The proposed voyage would continuously circle the globe, covering most of the world's coastal regions. Its large fleet of commuter aircraft and hydrofoils would ferry residents and visitors to and from shore.

The program has also said that the propellers would be a series of 400 fully-rotational azipods; despite the high number of screws, the ship would still be the slowest in the world. Despite an initially stated in-service date of 2001, construction has not even begun as of 2008.
Net price estimates for the ship have risen from 6 billion US$ in 1999 to 11 billion US$ in 2002.

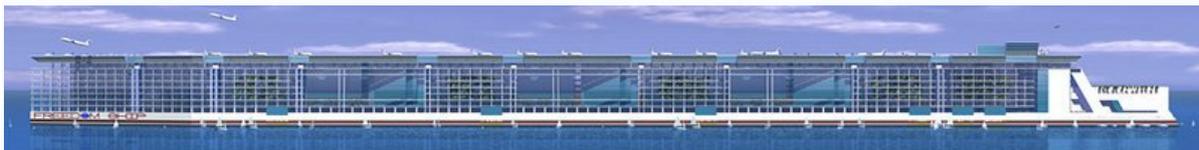

A side view of the proposed Freedom Ship. The largest existing ship in the world, the Knock Nevis, is approximately one third of this length.

The Freedom Ship has little in common with a conventional ship; it is actually nothing more than a big barge. The bolt-up construction and the unusually large amount of steel incorporated into the ship meets the design engineer's requirements for stability and structural integrity and the cost engineers requirements of "economic feasibility" but the downside is a severe reduction in top speed, making the ship useless for any existing requirements. For example, it would be too slow to be a cruise ship or a cargo ship.

But what if this big, overweight, barge was assigned a voyage that required slowly cruising around the world, hugging the shoreline, and completing one revolution every 3 years? If the designers then incorporated the following amenities into this barge, what would be the results?

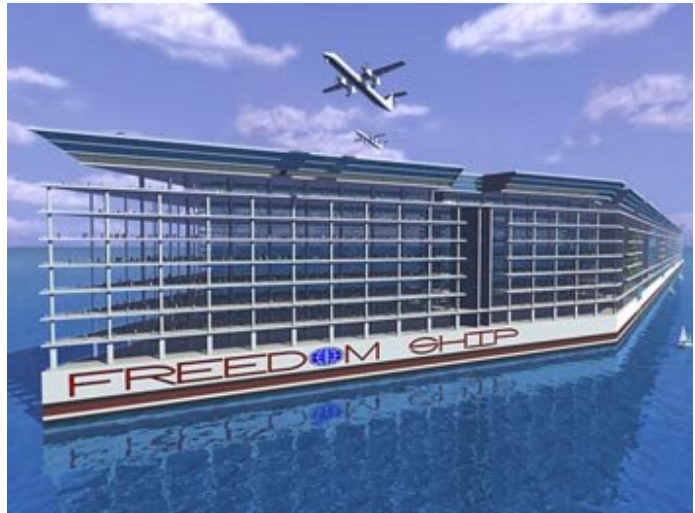
Freedom Ship (Front view).

- 18,000 living units, with prices in the range of $180,000 to $2.5 million, including a small number of premium suites currently priced up to $44 million.

- 3,000 commercial units in a similar price range
- 2,400 time-share units
- 10,000 hotel units
- A World Class Casino
- A ferryboat transportation system that provides departures every 15 minutes, 24 hours a day, to 3 or more local cities giving ship residents access to the local neighborhood and up to 30,000 land-based residents a chance to spend a day on the ship.
- A World-Class Medical Facility practicing Western and Eastern medicine as well as preventive and anti-aging medicine.
- A School System that gives the students a chance to take a field trip into a different Country each week for academic purposes or to compete with local schools in numerous sporting events. For example; The Freedom Ship High School Soccer team plays a Paris High School team this week at home and an Italian team next week in Italy, while the Freedom Ship High School Band presents a New Orleans Jazz musical at a concert hall in London.
- An International Trade Center that gives on-board companies and shops the opportunity to show and sell their products in a different Country each week.
- More than 100 acres of outdoor Park, Recreation, Exercise and Community space for the enjoyment of residents and visitors.

**Project Habakkuk** or **Habbakuk** (spelling varies) was a plan by the British in World War II to construct an aircraft carrier out of Pykrete (a mixture of wood pulp and ice), for use against German U-boats in the mid-Atlantic, which was out of range of land-based planes.

The *Habakkuk*, as proposed to Winston Churchill by Lord Mountbatten and Geoffrey Pyke in December 1942, was to be approximately 2,000 feet (610 m) long and 300 feet (91 m) wide, with a deck-to-keel depth of 200 feet (61 m), and walls 40 feet (12 m) thick. It was to have a draft of 150 feet, and a displacement of 2,000,000 tons or more, to be constructed in Canada from 280,000 blocks of ice. The ice *Habakkuk* itself was never begun.

**Artic and Antarctic (Southern) oceans**. The ice fields of these oceans will be used for getting float platforms.

The amount of sea ice around the poles in winter varies from the Antarctic with 18,000,000 km² to the Arctic with 15,000,000 km². The amount melted each summer is affected by the different environments: the cold Antarctic pole is over land, which is bordered by sea ice in the freely-circulating Southern Ocean.

The Arctic Ocean occupies a roughly circular basin and covers an area of about 14,056,000 km² (5,440,000 sq mi). The situation in the Arctic is very different from Antactic sea (a polar sea surrounded by land, as opposed to a polar continent surrounded by sea) and the seasonal variation much less, consequently much Arctic sea ice is multi-year ice, and thicker: up to 3–4 meters thick over large areas, with ridges up to 35 meters thick. An **ice floe** is a floating chunk of sea ice that is less than 10 kilometers (six miles) in its greatest dimension. Wider chunks of ice are called **ice fields**.

The North Pole is significantly warmer than the South Pole because it lies at sea level in the middle of an ocean (which acts as a reservoir of heat), rather than at altitude in a continental land mass. Winter (January) temperatures at the North Pole can range from about −43 °C (−45 °F) to −26 °C (−15 °F), perhaps averaging around −34 °C (−30 °F). Summer temperatures (June, July and August) average around the freezing point (0 °C, 32 °F). In midsummer of South pole, as the sun reaches its maximum elevation of about 23.5 degrees, temperatures at the South Pole average around −25 °C (−12 °F). As the six-month 'day' wears on and the sun gets lower, temperatures drop as well, with temperatures around sunset (late March) and sunrise (late September) being about −45 °C (−49 °F). In winter, the temperature remains steady at around −65 °C (−85 °F). The highest temperature ever recorded at the Amundsen-Scott South Pole Station is −13.6 °C (7.5 °F), and the lowest is −82.8 °C (−117.0 °F) (however, this is not the lowest recorded anywhere on earth, that being −89.6 °C (−129.28 °F) at Vostok Station) on July 21, 1983.

## Descriptions and Innovations

The author's idea is to use a cheap floating platform taken from the ice fields in Arctic and Antarctic oceans for the floating cities, island, and states. These cheap platforms protected by air-film (bottom and sides) and conventional insulating cover (top) and having cooling systems to deal with any leak-through heating can sustain the platform for an unlimited time. They can be increased in number or size at any time, float in warm oceans, travel to different continents and countries, serve as artificial airports, harbors and other marine improvements, as well as floating cities and industrial bases for virtually any use.

One possible means of construction is as follows: A scouting aircraft (helicopter) confirms a satellite-surveyed ice field as suitable and delives to it a small tractor with extensible wire–saw. The tractor saws up the ice platform to hew there from a platform of a specified size (including allowance for melting before insulation for example, 500×500×10 m)(fig.4a) and an ice-braker ship tows this platform to open water. Here the platform is equipped with air-film covers, protected by from warm water on all sides. Platform is towed to a place where it will be provided for with final protection and other improvements; a suitable location for building the city or other floating improvement that it will come. One method of adding thermal protection of the ice is the following: The double film is submerged lower than the bottom of platform, moved under the platform (or the platform is moved over film) and filled with air. The air increases the lift force of platform and protects the bottom, sides and top of the platform from contact with warm water and air (fig.4b, pointer 3). Simultaneously, the coolant fluid (it may be chilled air) flows throgh the cooling tubes 4 (fig.4b) and keeps the ice at lower than melting, or indeed softening, point.

The top side of the platform may be covered with conventional heat protection and insulation means on top of which construction elements may be added. (film, ground, asphalt, concrete plates, houses, buildings, gardens, airdrome runways, and so on).

The other method allows us to custom-produce ice of any thickness and composition, including ices of low density (high lift force). Thin plastic tubes are located under the ice-bottom to be (which may be isolated from circulation by a film barrier) and and cold air (in the polar regions, or in winter, simple outdoor air) is blown throgh them. One freezes the water and produces an ice platform. The ice has a lower density and a high lift force (load capability) because the ice has internal channels (tubes) filled by air. We may evade spending energy for it in cold countries or in winter. The arctic (antarctic) winter air has temperature up to $-(40$ -$50)^{\circ}$C. In the Arctic, Ocean water is useful as a heat source (having $0^{\circ}$C) which can heat the outer air up to -3-5$^{\circ}$C, turning an the air turbine, the turbine then turning the pump air ventilator. The corresponding estimation is in theoretical section. We can get the ice density of $\gamma$ = 500 kg/m$^3$ having load capability of 500 kg/m$^3$ (the conventional ice has the lift force 80 kg/m$^3$). For decreasing the ice density may be used cork material or other low density fillers..

In second method we can produce platform from **Pykrete** (also known as **picolite**). That is a composite material made of approximately 14% sawdust (or, less frequently, wood pulp) and 86% water by weight then frozen, invented by Max Perutz. Pykrete has some interesting properties, notably its relatively slow melting rate (due to low thermal conductivity), and its vastly improved strength and toughness over pure ice, actually closer to concrete, while still being able to float on water. Pykrete is slightly harder to form than concrete, as it expands while freezing, but can be repaired and maintained from the sea's most abundant raw material.

The pykrete properties may be significantle impruved by using the cheap artificial strong fibers as basalt fibers, class or mineral wool, and others.

The composites made by mixing cork granules and cement have low thermal conductivity, low density and good energy absorption. Some of the property ranges of the composites are density (400–1500 kg/m³), compressive strength (1–26 MPa) and flexural strength (0.5–4.0 MPa).

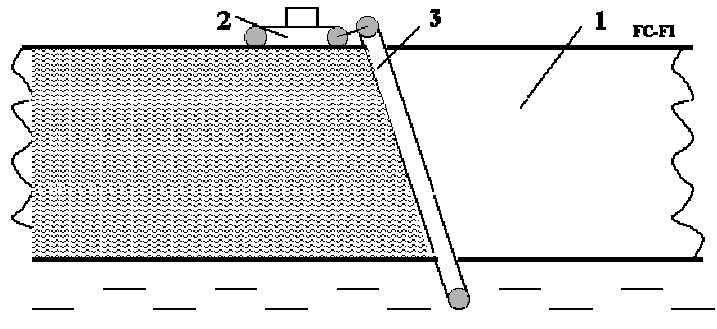

**Fig.3**. Cutting of floating platform from ice field. *Notations*: 1 – ice field in arctic (Antarctic) ocean; 2 – small tractor with band-saw or wiresaw; 3 – mechanical band saw or wiresaw.

The platform of floating city has protection (walls) 6 (fig.5) against storm ocean waves, joints 7 (fig.5b) which decrease the platform stress in storms, azimuth thruster propellers for maneuvering and moving. The platform may also have an over film dome (fig.5b, pointer 9) as it offered in [2. 11]. This dome creates a warm constant "deck" temperature, protects the floating city from strong winds and storms.

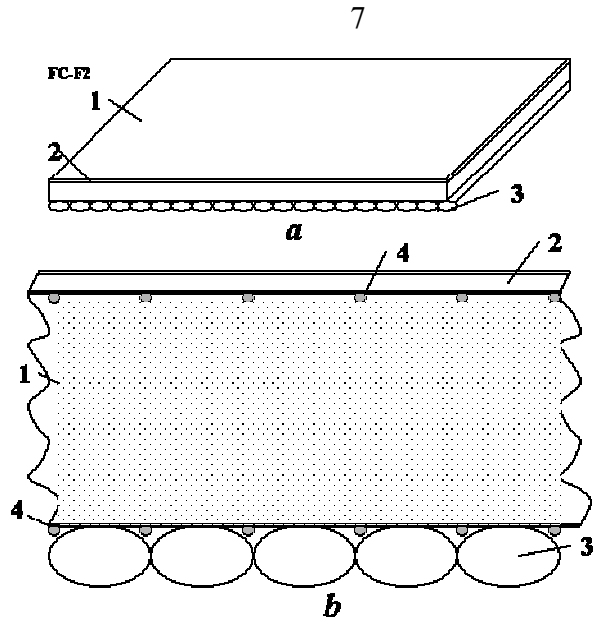

**Fig.4**. Ice platform prepared for floating city. (a) Common view, (b) Cross-section of platform. *Notations*: 1 – ice; 2 – top heat protection; 3 – low (bottom) heat protection and floating support (inflatable air balloon); 4 – cooling tubes.

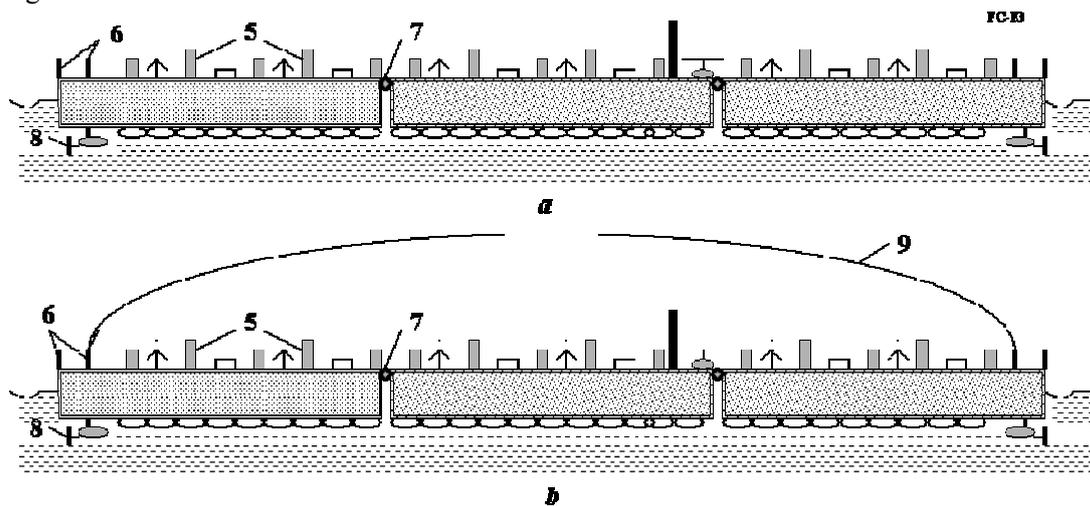

**Fig.5**. Floating city on ice platform: (a) Open floating city, (b) Floating city closed by film. *Notations*: 5 – city; 6 – protection from ocean waves in storm; 7 – turning connection (joint) of separated ice platform; 8 – fully-rotation azimuth thruster propellers; 9 – film dome.

Summary of innovations:
1) Using a big natural ice platform for building large floating cities, islands and states.
2) Technology for getting these platforms from the natural ice fields (saw up).
3) Technology (artificial freezing without spending energy) for getting the artificial high lift force ice platform of any thickness (that means any load capability) from low density ice.
4) Composite material where ice is a matrix (base) and cork (or other material) as stuff.
5) Heat protection for natural ice fields by air film balloons.
6) Building the platforms from separated ice segments and connection them by joints.
7) It is offered the protection of the suggested platform by special double walls 6 (fig.5) from ocean storm waves.

8) Protection of the suggested platform by the special transparent film 9 (dome) (fig.5b) and creating a constant temperature in the floating city, plus protection from strong winds and storm.

## Theory of estimation and computation

**1. Material**. The important values and characteristics of candidate materials for floating platforms are their price, lift force (in water), life time, strength, chemical stability in water, reliability, and so on. The water lift force of matter ($L_f$) is difference between density of water ($d_w$) and density of platform matter ($d_m$).

$$L_f = d_w - d_m, \qquad (1)$$

where all value are in kg/m$^3$.

Air is the cheapest material, having the most lift force. However it needs a strong cover (in vessels, balloon, film) which can significantly increase the cost of the installation. The other lack (disadvantage) of using air is loss of lift force in case of damage to its' container.

Ice is a cheap substance. It may be mined, rather than of necessity built, into a ready floating platform. But it has a small buoyancy force and low melting temperature, which is lower then the temperature of ocean water. We can decrease these disadvantages by using special air balloons under the platform, heat protection materials and barriers and a refreezing system. If we initially produce the platform by custom freezing, we can produce the custom-tailored, strong light ice having a high lift (buoyancy) force.

The other systems are metal or concrete constructions, filled by rock, foam plastic, aerocrete and so on. Their disadvantages are well known: A high cost and a huge procurement necessary for big installations (platforms).

Some materials and their properties are presented in Tables 1 and 2.

**2. Computation of heat protection**. We use in our project the cheap natural ice platform variant.

The heat loss flow per 1 m$^2$ by convection and heat conduction is (see [13]):

$$q = k(t_1 - t_2), \quad \text{where} \quad k = \frac{1}{1/\alpha_1 + \sum_i \delta_i / \lambda_i + 1/\alpha_2}, \qquad (2)$$

where $k$ is heat transfer coefficient, W/m$^2$·K; $t_{1,2}$ are temperatures of the inter and outer multi-layers of the heat insulators, °C; $\alpha_{1,2}$ are convention coefficients of the inter and outer multi-layers of heat insulators ($\alpha = 30 \div 100$), W/m$^2$K; $\delta_i$ are thickness of insulator layers; $\lambda_i$ are coefficients of heat transfer of insulator layers (see Table 1), m. The magnitudes $\alpha$ are:
1. From water to metal wall $\alpha = 5000$ W/m$^2$K.
2. From gas to wall $\alpha = 100$ W/m$^2$K.
3. From heat isolator to air $\alpha = 10$ W/m$^2$K.

For example, let us estimate the heat flow from water to the bottom surface of ice platform protected by the small air balloons.

Assume the average thickness of air balloons is $\delta = 1$ m, the temperature of water at depth 10 m is 10°C. The heat flow from water to ice platform is

$$k \approx \frac{\lambda}{\delta} = \frac{0.0244}{1} = 0.0244 \; \frac{\text{W}}{\text{m}^2\text{K}};$$

$$q = k(t_2 - t_1) = 0.0244(10 - 0) = 0.244 \; \frac{\text{W}}{\text{m}^2\text{K}}.$$

That ($0.244 \; \frac{\text{W}}{\text{m}^2\text{K}}$) is a small value. This heat must be deleted by cooling liquids or fluids (ie cooled, force-driven air circulated for that purpose).

Estimate now the heat flow from outer air to the top platform surface protected by 0.1 m wood gasket and 0.4 m humid soil. The air temperature is 25°C.

$$k \approx \frac{1}{\delta_1/\lambda_1 + \delta_2/\lambda_2} = \frac{1}{0.1/0.2 + 0.4/0.657} = 0.9,$$

$$q = k(t_2 - t_1) = 0.9 \cdot 25 = 22.5 \ \frac{W}{m^2 K}.$$

If we change the wooden gasket for an asbestos plate of same thickness the heat flow decreases to q = 17 W/m²K. Places where are houses, buildings and other constructions having concrete base will have q = 10 – 15 W/m²K. Using the wools or air protection significantly decreases the head loss through top platform surface. The average heat loss of top platform surface is about 15 W/m²K. If we insert the air black gap 15 – 25 cm, this heat loss decreases to 1 – 2 W/m²K. The side part of the floating platform may be protected the same as the bottom surface.

**3. Freezing of platform.** The freezing of 1 kg water requests energy

$$Q = c_p(t_2 - t_1) + \lambda_p \approx \lambda_p, \qquad (3)$$

where $c_p$ = 4.19 kJ/kg·C is energy needed for cooling water in 1°C; $\lambda_p$ = 334 kJ/kg is energy needed for freezing 1 kg of water.

The energy needed for freezing may be received from the cold arctic air. That computed by equation

$$Q = c_{p,a}(t_2 - t_1), \qquad (4)$$

Here $c_{p,a}$ = 1 kJ/kg·K for air.

The computation shows for freezing 1 kg water we need about 22 kg air in temperature -20 C. Every 1 kg air heated from -20 C to -5 C in ocean water absorbs about 15 kJ/kg in heat energy.

**4. Other heat flows.** The radiation heat flow per 1 m²s of the service area computed by equations (5):

$$q = C_r\left[\left(\frac{T_1}{100}\right)^4 - \left(\frac{T_2}{100}\right)^4\right], \quad \text{where} \quad C_r = \frac{c_s}{1/\varepsilon_1 + 1/\varepsilon_2 - 1}, \quad c_s = 5.67 \ [W/m^2 K^4], \quad (5)$$

where $C_r$ is general radiation coefficient, $\varepsilon$ are black body rate (Emittance) of plates; $T$ is temperatures of plates, °K.

The radiation flow across a set of the heat reflector plates is computed by equation

$$q = 0.5\frac{C'_r}{C_r}q_r, \qquad (6)$$

where $C'_r$ is computed by equation (5) between plate and reflector.

The data of some construction materials is found in Table 1, 2.

Table 1. [13], p.351.[14], p.73. Heat Transfer

| Material | Density kg/m³ | Heat transfer λ=W/m°C | Heat capacity kJ/kg°C |
|---|---|---|---|
| Concrete | 2300 | 1.279 | 1.13 |
| Baked brick | 1800 | 0.758 | 0.879 |
| Ice | 920 | 2.25 | 2.26 |
| Show | 560 | 0.465 | 2.09 |
| Glass | 2500 | 0.744 | 0.67 |
| Steel | 7900 | 45 | 0.461 |
| Air | 1.225 | 0.0244 | 1 |
| Asphalt | 2110 | 0.6978 | 2.09 |
| Asbestos plate | 770 | 0.1162 | 0.810 |

| Oak | 800 | 0.207 | 1.758 |
|---|---|---|---|
| Humid soil | 1700 | 0.657 | 2.01 |
| Mineral wool | 200 | 0.0465 | 0.921 |
| Dry sand | 1500 | 0.326 | 0.795 |
| Glass wool | 200 | 0.037 | 0.67 |
| Slag wool | 250 | 0.0698 | - |
| Aluminum | 2670 | 204 | 0.921 |
| Water | 1000 | 0.5513 | 4.212 |
| Sold rubber | 1200 | 0.169 | 1.382 |
| Aerocrete | - | 0.07-0.32 | - |
| Foam plastic | - | 0.043-0.058 | - |
| Reinforced concrete | - | 1.55 | - |

As the reader sees, the air layer is the best heat insulator. We do not limit its thickness $\delta$.

The thickness of the dome envelope, its sheltering shell of film, is computed by formulas (from equation for tensile strength):

$$\delta_1 = \frac{Rp}{2\sigma}, \quad \delta_2 = \frac{Rp}{\sigma}, \tag{7}$$

where $\delta_1$ is the film thickness for a spherical dome, m; $\delta_2$ is the film thickness for a cylindrical dome, m; $R$ is radius of dome, m; $p$ is additional pressure into the dome, N/m$^2$; $\sigma$ is safety tensile stress of film, N/m$^2$.

For example, compute the film thickness for dome having radius $R = 100$ m, additional air pressure $p = 0.01$ atm ($p = 1000$ N/m$^2$), safety tensile stress $\sigma = 50$ kg/mm$^2$ ($\sigma = 5 \times 10^8$ N/m$^2$), cylindrical dome.

$$\delta = \frac{100 \times 1000}{5 \times 10^8} = 0.0002 \, m = 0.2 \, mm \tag{8}$$

The dynamic pressure from wind is

$$p_w = \frac{\rho V^2}{2}, \tag{9}$$

where $\rho = 1.225$ kg/m$^3$ is air density; $V$ is wind speed, m/s.

For example, a storm wind with speed $V = 20$ m/s (72 km/h), standard air density is $\rho = 1.225$ kg/m$^3$. Then dynamic pressure is $p_w = 245$ N/m$^2$. That is four time less than internal pressure $p = 1000$ N/m$^2$. When the need arises, sometimes the internal pressure can be voluntarily decreased, bled off.

**Table 2.** Material properties

| Material | Tensile strength kg/mm$^2$ | Density g/cm$^3$ | **Fibers** | Tensile strength kg/mm$^2$ | Density g/cm$^3$ |
|---|---|---|---|---|---|
| **Whiskers** | | | | | |
| AlB$_{12}$ | 2650 | 2.6 | QC-8805 | 620 | 1.95 |
| B | 2500 | 2.3 | TM9 | 600 | 1.79 |
| B$_4$C | 2800 | 2.5 | Allien 1 | 580 | 1.56 |
| TiB$_2$ | 3370 | 4.5 | Allien 2 | 300 | 0.97 |
| SiC | 1380-4140 | 3.22 | Kevlar or Twaron | 362 | 1.44 |
| **Material** | | | Dynecta or Spectra | 230-350 | 0.97 |
| Steel prestressing strands | 186 | 7.8 | Vectran | 283-334 | 0.97 |
| Steel Piano wire | 220-248 | | E-Glass | 347 | 2.57 |
| Steel A514 | 76 | 7.8 | S-Glass | 471 | 2.48 |
| Aluminum alloy | 45.5 | 2.7 | Basalt fiber | 484 | 2.7 |
| Titanium alloy | 90 | 4.51 | Carbon fiber | 565 | 1,75 |
| Polypropylene | 2-8 | 0.91 | Carbon nanotubes | 6200 | 1.34 |

## Projects

The estimation of different variants of floating platforms is presented below in Table 3.

**Table 3.** Estimation of different variants of floating platforms

| # | Type of floating platform | Height, m | Cost,* $/m$^2$ | Life time, Year | Load capasity, ton/m$^2$ | Main-tains, $/m$^2$year | Draught m | Mass of platform, ton/m$^2$ | Cooling Energy, W/m$^2$ |
|---|---|---|---|---|---|---|---|---|---|
| 1 | Air-steel cylinder with steel walls | 10<br>20 | 100<br>200 | 30-50 | 7<br>17 | 1<br>2 | 7.6<br>18.2 | 0.6<br>1.2 | 0 |
| 2 | Steel cubs with net walls and air balloons | 10<br>20 | 150<br>300 | 40-60 | 7<br>17 | 2<br>4 | 7.5<br>18 | 0.5<br>1 | 0 |
| 3 | Steel cubs with net walls and foam plastic filler | 10<br>20 | 150<br>130 | 40-60 | 7<br>17 | 1<br>2 | 7.6<br>18.2 | 0.6<br>1.2 | 0 |
| 4 | Concrete empty cub with walls 0.1m, 100 $/ton | 5<br>10 | 400<br>800 | 100-200 | 2.4<br>6 | 0.5<br>1 | 4<br>9.2 | 1.6<br>3.2 | 0 |
| 5 | Aero Crete $\gamma = 500$ kg/m$^3$ | 10<br>20 | 220<br>440 | 100-200 | 4<br>7 | 1<br>2 | 9<br>17 | 6<br>17 | 0 |
| 6 | Ice and 1 m air heat protection in bottom | 5<br>10 | 2<br>3 | ∞ | 1<br>2 | 4<br>6 | 5<br>10 | 4<br>8 | 2 W/m$^2$<br>2 W/m$^2$ |
| 7 | Air ice, $\gamma = 500$ kg/m$^3$ and 1 m air heat protect. | 20<br>30 | 4<br>6 | ∞ | 9<br>15 | 5<br>10 | 19<br>30 | 10<br>15 | 2 W/m$^2$<br>2 W/m$^2$ |

* only material.

The estimation cost of 1 m$^2$ of the platform in the contemplated "Freedom Ship" (the cost of cabins are included) is $33,100/m$^2$ (2002). At the present time (2008) this cost has increased by a factor of two times more. Average cost of 1 m$^2$ of apartment in many cities is about $1000/m$^2$ (USD).

## Discussing

Advantages and disadvantages of the offered method.
Advantages:
1. The offered method is cheapest by tens-to-hundreds of times relative to conventional shipbuilding, and beats nearly all but the remotest and most valueless land for cheapness as a construction substrate—yet the product may be relocated to within meters from some of the most valuable real estate on Earth—i.e. docked near Tokyo or Manhattan or Shanghai.
2. Unlimited increase of useful area possible.
3. Easy increasing of load capacity by additional freezing of new platform bottom area; and ease of restoring a damage sector.
4. High security.
5. Unlimited life time.

Disadvantages:
1. Need for a permanent but small energy expenditure (in warm climates) for maintaining the ice at freezing temperatures.

## Results

It is a promising new method for obtaining a cheap ice platform suitable for many profitable engineering purposes, and for colonization the World Ocean.

## Acknowledgement

The author wishes to acknowledge Joseph Friedlander for correcting the author's English and useful suggestions and technical advice.

**References**
(The reader find some author's works in http://Bolonkin.narod.ru/p65.htm, http://Arxiv.org  Search: Bolonkin and books: Bolonkin A.A., Non-Rocket Space Launch and flight:, Elsevier, 2006, 488 pgs., Bolonkin A.A., "New Concepts, ideas, and Innovations in Technology and Human life", NOVA, 2008, 400 pg.)

1. Bolonkin, A.A., Irrigation without water (closed-loop water cycle). http://arxiv.org search "Bolonkin"(2007).
2. Bolonkin, A.A. and R.B. Cathcart, Inflatable 'Evergreen' Dome Settlements for Earth's Polar Regions. Clean. Techn. Environ. Policy. DOI 10.1007/s10098.006-0073.4 .
3. Bolonkin A.A., Control of Regional and Global Weather. 2006. http://arxiv.org search for "Bolonkin".
4. Bolonkin A.A., Cheap Textile Dam Protection of Seaport Cities against Hurricane Storm Surge Waves, Tsunamis, and Other Weather-Related Floods, 2006. http://arxiv.org.
5. Bolonkin, A.A. and R.B. Cathcart, Antarctica: A Southern Hemisphere Windpower Station? Arxiv, 2007.
6. Cathcart R.B. and Bolonkin, A.A. Ocean Terracing, 2006. http://arxiv.org.
7. Bolonkin, A.A. and R.B. Cathcart, The Java-Sumatra Aerial Mega-Tramway, 2006. http://arxiv.org.
8. Bolonkin, A.A., "Optimal Inflatable Space Towers with 3-100 km Height", *Journal of the British Interplanetary  Society* Vol. 56, pp. 87 - 97, 2003.
9. Bolonkin A.A., *Non-Rocket Space Launch and Flight, Elsevier*, London, 2006, 488 ps.
10. Macro-Engineering: *A Challenge for the Future*. Springer, 2006. 318 ps. Collection articles.
11. Bolonkin A.A., Cathcart R.B., Inflatable 'Evergreen' for Polar Zone Dome (EPZD) Settlements, 2006. http://arxiv.org search "Bolonkin".
12. Bolonkin A.A., Inflatable Dome for Moon, Mars, Asteroids and Satellites**,** Presented as paper AIAA-2007-6262 by AIAA Conference "Space-2007", 18-20 September 2007, Long Beach. CA, USA.
13. Naschekin, V.V., Technical thermodynamic and heat transmission, Public House of High University, Moscow, 1969 (in Russian).
14. Koshkin H.I., Shirkevich M.G., Directory of elementary physics, Moscow, Nauka, 1982.
15. Wikipedia. Some background material in this article is gathered from Wikipedia under the Creative Commons license.